\documentclass[12pt]{article}
\usepackage[a4paper, margin=1in]{geometry}
\usepackage{amsmath, amssymb}
\usepackage{authblk}
\usepackage{graphicx}
\usepackage{cite}
\usepackage{hyperref}
\usepackage{subcaption}
\usepackage{float}
\usepackage{mathrsfs}
\usepackage[margin=1in]{geometry} % Optional: adjust page margins
\usepackage[section]{placeins}
\usepackage{orcidlink}
\usepackage{epstopdf}

\hypersetup{colorlinks=true, linkcolor=blue, citecolor=blue, urlcolor=blue}

\title{\bf Unified Cosmological Scenario in Holographic $f(Q)$ gravity: From Inflation to Late-Time Acceleration}

\author[1]{Moli Ghosh \,\orcidlink{0009-0003-1025-5832} \thanks{molig2018@gmail.com; moli.ghosh@s.amity.edu}}
\author[2]{Can Aktaş \,\orcidlink{0000-0002-0603-7862}\thanks{canaktas@comu.edu.tr}}
\author[3]{ Surajit Chattopadhyay \,\orcidlink{0000-0002-5175-2873}\footnote{Corresponding author}\thanks{schattopadhyay1@kol.amity.edu; surajitchatto@outlook.com
}}
\affil[1,3]{\small Department of Mathematics, Amity University Kolkata, Major Arterial Road, Action Area II, Rajarhat, Newtown, Kolkata 700135, India.}
\affil[1] {\small Department of Mathematics, Mrinalini Datta Mahavidyapith, Kolkata-700051, India.}
\affil[2]{Department of Mathematics, Faculty of Science , Çanakkale Onsekiz Mart University, Terzio\u{g}lu Campus, 17100 Çanakkale, Turkey.}

\date{}
\begin{document}

\maketitle
\begin{abstract}
\noindent
The present paper reports a study of a unified cosmological scenario in the framework of holographic $f(Q)$ gravity, where, in a single theoretical setup, both the early inflationary epoch and the late-time accelerated epoch are studied. Considering $f(Q)=\zeta Q^n$, we reconstruct the Hubble parameter in the presence of Barrow holographic fluid and study the inflationary behaviour through the slow-roll parameters, scalar spectral index $n_s$, and tensor-to-scalar ratio $r$. The obtained inflationary predictions are found to be consistent with the latest Planck 2018 observational constraints, with a very small value of the tensor-to-scalar ratio. In the next phase we extend the study by including the matter sector. The Chevallier-Polarski-Linder (CPL) parametrization is used to connect the theoretical model with observational cosmology. Using combined Cosmic Chronometer (CC) and Baryon Acoustic Oscillation (BAO) datasets, the study constrains the model parameters through Markov Chain Monte Carlo (MCMC) analysis. From the observational results obtained this way, the study concludes that at low redshifts, the  holographic $f(Q)$ model considered here remains compatible with the standard $\Lambda$CDM and mild deviations are observed at higher redshift. We have also done the AIC and BIC analysis for our analysis and commented about the goodness of the fit against $\Lambda$CDM. Hence, the present framework provides a viable unified description of inflation and late-time cosmic acceleration within holographic $f(Q)$ gravity.
\end{abstract}
\clearpage
\tableofcontents

\section{Introduction}  
\label{sec:intro}

One of the most fundamental issue in modern theoretical cosmology is whether the universe began with singularity or through a non-singular process. Naturally , the initial singularity of Big-Bang \cite{ar1,ar2} is more convenient description as we can imagine a zero-sized, infinitely dense, and hot Universe. However we can not exclude the cyclic cosmological evolution in which the  Universe shrinks to zero. In the last few decades the cosmologist community experience with great surprise that our Universe is expanding in an accelerated way. The period of late time accelerating area is reffered as dark energy era \cite{Caldwell2009,Caldwell2009Nature,Li2011,Bamba2012,Lobo2008,Gubitosi2013,Bloomfield2013,Gleyzes2013,Li2013} and several models have been proposed to explain it. Some of them use  scalar fields particularly quintessence \cite{Olivares2005, Capozziello2002,Sahni2002, Capozziello2003, Carroll1998, Wang1998}  while others explore modified gravity in different forms. 

In addition to the late time acceleration,the Univserse also experienced another acceleration era during its first stages of evolution which is known as inflation \cite{Linde2008,Lyth1999,Linde1983,Bamba2015,Baumann2009,Martin2014}  and the latest observational data have actually constrained this era \cite{Planck2016, BICEP2016}. From a theoretical perspective, the inflationary model of the early universe successfully resolved several short comings of the standard Big Bang cosmology, such as the horizon problem, monopole problem, etc. However any plausible inflationary scenario must be alligned with observational constraint especially those derived from the Planck Collaboration data, which impose stringent limits on the dynamics of the inflationary epoch \cite{Planck2016}. The findings from Planck's mission have ruled out a significant number of previously proposed inflationary models, establishing these observations as a crucial benchmark for assessing the viability of any inflationary framework. Consequently compatibility with observational data is considered as an essential requirement for any model to be considered as a candidate for inflation. 

There are several avenues through which the unified description of inflation and dark energy has been proposed. Such as in the context of modified gravity \cite{Nojiri2003}  and of string theory \cite{Liddle2006}, where the IR description of string theory serves both purposes. The unification using the holographic approach has been explored in \cite{Nojirigrg2006, Li2008}. There is also a proposal for dynamically evolving dark energy \cite{Xia2008}, which unifies dark energy and inflation in a very natural way. This has also been addressed in the context of modified $F(R)$ Horava-Lifshitz gravity in \cite{Nojiri2010}. Apart from that, these issues have been addressed in other paradigms \cite{Koivisto2009,Xia2008,Liddle2008,Rinaldi2015,Nojiri2016,Cai2008,Geng2011,Vacaru2015}. Even though modified gravity is a very lucrative avenue to explore, it has its own problems, such as the violation of the principle of equivalence \cite{Olmo2007}, non-standard energy conditions, and non-canonical thermodynamic interpretation \cite{Bamba2016,Zubair2016}, just to name a few. Since Einstein's gravity is extremely well tested by observations, any modified gravity also has strict phenomenological bounds set by experiments such as the solar system test \cite{Nojiri2007, Olmo2005,Faraoni2006, Iorio2011,Allemandi2007,Zhang2007} . The more general study of the $f(R)$ limit, which is consistent with the Newtonian gravity limit, has been studied in \cite{Capozziello2007}. There are many studies on the stability of $f(R)$ gravity, as well as on the parameter constraints from the cosmic chronometers (CC) data set. As the landscape of the modified gravity is far more richer than the Einstein gravity, one can find the detailed discussion about how to construct such models in various references such as \cite{Faraoni2005,Carloni2017,Duchaniya2025,Lohakare2025,Narawade2024}.

A well-known alternate theory for understanding dark energy,  is based on the holographic principle, which was proposed by Hooft \cite{refs1}. Cosmologist were confronted with two fundamental problem from the discovery of dark energy: one is the fine-tuning problem, and another is the coincidence problem. Holographic principle may provide some clue for solving these issues.  According to Cohen et al. introduced ``UV-IR" relationship highlighting the effective quantum field theory, a system of size L has its entropy and energy confined by the Bekenstein entropy bound and black hole mass, respectively. This suggests that low-energy physics outside of black holes can only be described by quantum field theory \cite{refs2}. The key step  in the application of holograpic
principle at cosmological framework is that the entropy of Universe horizon  is proportion to is proportional
to its area, similarly to the Bekenstein-Hawking entropy
of a black hole. HDE has been extensively studied in both its basic and expanded variants, and throughout time, the model has shown considerable success \cite{Saridakis2020,Shekh2021,Kalkan2026}.
The interoperability of the HDE models with observational data has been one of their main achievements \cite{Li2009,Fayaz2013}.

%Following that, Li et al. suggested that the event horizon's size serves as the infrared cut-off important to the dark energy. The dark energy density they found can be expressed as $\rho_{de} = 3c^2{M_{p}}^2{R_{h}}^2$ , where $R_h$ is our universe's future horizon \cite{Li2004}.

Modified gravity theories offer an important motivated approach for a better understanding of the evolution of the Universe, encompassing both the early inflationary phase and the late-time accelerated expansion linked to dark energy \cite{CANTATA2021}. In GR, the Levi-Civita connection is required to describe its gravitational interaction. The decision is predicted on the hypothesis of free geometry of torsion and nonmetricity. In addition we have to remember that the general affine connection has a more generic expression \cite{Hehl1995}. Among different modified gravity symmetric teleparallel gravity provide a distinct description in which non-metricity scalar $Q$ is the source of gravitational interaction. In $f(Q)$ gravity nonmetricity remain non zero while curveture and torsion are bound to disappear.  The
symmetric teleparallel gravity or $f(Q)$ gravity, hosted by Jimenez et al. \cite{Jimenez2018} has attracted a lot of attraction  lately as a feasible substitute for conventional cosmology and as a potential route toward novel gravitational physics outside the $\Lambda CDM$ paradigm.

Although the existing literatures indicate a handful of works on holographic dark energy and $f(Q)$ gravity separately, unified treatment of early-time inflation and late-time cosmic acceleration within a single holographic $f(Q)$ framework has received comparatively less attention. In the present work, we consider a power-law form $f(Q)=\zeta Q^n$ and reconstruct the Hubble parameter in the presence of holographic fluid to study inflationary dynamics through the observables $n_s$ and $r$. In addition, the late-time cosmological behaviour is examined using CPL parametrization together with CC+BAO observational datasets through MCMC analysis. Therefore, the novelty of the present study lies in the fact that the study intends to provide a single consistent holographic $f(Q)$ framework capable of describing both inflationary and late-time accelerated phases of the universe while remaining compatible with observational constraints. The rest of the paper is organized as follows. In Section \ref{sec:f(Q)}, the theoretical framework of $f(Q)$ gravity is reviewed briefly, and the holographic fluid is discussed in the context of symmetric teleparallel gravity. Section \ref{sec:infl} reports the reconstruction of the Hubble parameter in terms of the e-folding number and investigates the inflationary dynamics through the slow-roll parameters, scalar spectral index, and tensor-to-scalar ratio, followed by a comparison with observational constraints. Section \ref{sec:matter} studies late-time cosmology in the presence of a matter sector. In this Section, CPL parametrization is implemented, and the model parameters are constrained using CC+BAO observational datasets through MCMC analysis. Finally, in Section \ref{sec:conclusion}, we summarize the main results and discuss the cosmological implications of the present holographic $f(Q)$ gravity model.

  \section{Theoretical framework of $f(Q)$ gravity}
  \label{sec:f(Q)}
 
  \subsection{Basics of Symmetric Teleparallel Gravity}
We start by pointing out that a connection plays an important role in the transportation of tensors across a manifold.  The gravitational interaction in symmetric teleparallel gravity is fully encoded in the non-metricity scalar $Q$ \cite{Jimenez2018,Frusciante2021,DAmbrosio2022 }. As the theory is based on a flat, torsion-free link, torsion and curvature vanish in the same way. Consequently, the Levi-Civita connection in standard GR is replaced by a description of gravity based only on non-metricity.   
%{The authors Capozziello et al.\cite{revise1} looked into the preservation of information stored in bosons within the framework of $f(Q)$ gravity, driven by the need to preserve information and create alternatives to Einstein's gravity. By doing this, they used the Bogoliubov transformation to study the evolution of bosonic Gaussian states from the far past to the distant future. In the end, they concluded that equivalent particle productions occur even when one analyzes theories of symmetric teleparallel gravity, much like in general relativity. Convenient and mimetic $f(Q)$ gravities were formulated by Nojiri and Odintsov \cite{revise2} in terms of four scalar fields and the metric, where $Q$ is the non-metricity scalar. Consequently, the well-definedness of the resulting field equations is demonstrated. They showed how to recreate the f(Q) models that realize any given FLRW spacetime by utilizing the field equations. To explain and even unite the inflation and dark energy epochs, they devised a handy form of $f(Q)$ gravity. In this context, let us mention that in this work, we have looked into both bouncing and inflationary scenarios. The exponential, matter, symmetric, and super-bounce models are the four bouncing models that we plan to investigate. To further investigate if kinetic energy is dominated and whether all of the bouncing models are consistent with inflationary conditions, we plan to use a scalar field in the inflationary scenario.}

In  metric tensor $g_{\alpha\beta}$ and connection term ${\Gamma}^{\nu}_{\alpha \beta}$ are treated independently $f(Q)$ gravity.  We define the connection of the non metricity  by 
\begin{equation}{\label{b1}}
    Q_{\lambda \alpha \beta}=  \nabla_{\lambda} g_{\alpha \beta}= \partial_{\lambda} g_{\alpha \beta}- {\Gamma}^{\lambda}_{\lambda \alpha } g_{\nu \beta } - {\Gamma}^{\lambda}_{\lambda \beta } g_{\nu \alpha }
\end{equation}
and corresponding traces are 
$Q_{\lambda}={{Q}^{\alpha}_{\lambda}}\;{\alpha}$ and $\tilde{Q_{\lambda}}= {Q}^{\alpha}_{\lambda \alpha}$.
The teleparallel equivalent of General Relativity (TEGR) uses  the Weitzenböck connection where torsion is the only quantity responsible for gravity and both curvature and non-metricity vanish identically. However, within the context of geometrodynamics a more general geometric structure may be taken into consideration. The metric-affine connection, which may be reffered as the most widespread connection—is decomposed into following independent components \cite{BeltranJimenez2019}

\begin{equation}{\label{b2}}
\tilde{\Gamma}^{\nu}_{\lambda\zeta}= {\Gamma}^{\nu}_{\lambda\zeta}+ {K}^{\nu}_{\lambda\zeta}+{L}^{\nu}_{\lambda\zeta}
\end{equation}
where ${\Gamma}^{\nu}_{\lambda\zeta}$ is the Levi-Civita connection, ${K}^{\nu}_{\lambda\zeta}$ is is the contorsion tensor and ${L}^{\nu}_{\lambda\zeta}$ is the disformation tensor. Moreover, the non-metricity conjugate is defined as 
\begin{equation}\label{b3}
  {P}^{\lambda}_{\beta\alpha}= -\frac{1}{4} {Q}^{\lambda}_{\beta\alpha}+ \frac{1}{2} {Q}^{\lambda}_{\beta}\;{\alpha}+ \frac{1}{4} (Q_{\zeta}-\tilde{Q_{\zeta}}) g_{\beta\alpha}- \frac{1}{4}
  \end{equation}
Now the  action of $f(Q)$ gravity is given by \cite{Jimenez2018, Jimenez2020}.

\begin{equation}\label{b4}
    S =  \int [-\frac{1}{16 \pi G} f(Q)+\mathcal{L}_{m}]\sqrt{-g} dx 
\end{equation}
 Varying the action Eq.(\ref{b4}) leads the field equation as\cite{Dialektopoulos2019,Jimenez2020}.

 \begin{equation}\label{b5}
-{T}_{\beta\alpha}= \frac{2}{\sqrt{-g}} \nabla_{\lambda}(\sqrt{-g} f_{Q} P^{\lambda}_{\beta\alpha})+ f \frac{1}{2} g_{\beta\alpha}+ f_{Q} (P_{\beta\gamma \delta} Q^{\gamma \delta}_{\alpha}- 2 Q_{\gamma \delta \beta} P^{\gamma \delta}_{\alpha})
 \end{equation}
 The energy momentum tensor defined by 
 \begin{equation}
     {T}_{\beta\alpha} = \frac{-2}{\sqrt{-g}} \frac{\delta(\sqrt{-g} \mathcal{L}_{m}) }{\delta g^{\beta\alpha}}
 \end{equation}

 On large scale approximation The Universe is assumed to be homogeneous and isotropic. Therefore we adopt FLRW metric and the scale factor is represented by $a(t)$. Now substituting the the FLRW metric in the above field equation  (Eq.(\ref{b5})), we obtain the field equation for $f(Q)$
 
 %We know that in large-scale approximation, our universe is homogeneous and isotropic. Therefore we consider the FLRW metric, and the scale factor is represented by $a(t)$. By putting the flat FLRW metric in the above field equation (Eq.(\ref{b5})), we get the following field equation for $f(Q)$ gravity 

\begin{equation}\label{b6}
    3H^2=\frac{1}{2f_{Q}}(-\rho +\frac{f}{2})
\end{equation}
\begin{equation}\label{b7}
    \dot{H}+3H^2+ \frac{\dot{f_{Q}}}{f_{Q}} H=\frac{1}{2f_{Q}}(p +\frac{f}{2})
\end{equation}
where $Q=6H^2$ is the non metricity tensor and $ f_{Q}=\frac{df}{dQ}$ \cite{Ghosh2024}\\
The power law form of $f(Q)$ gravity has attracted significant attention in modified gravity and cosmological investigations because  of its rich phenomenological behaviour. Moreover power law form are widely used in modified gravity because they can effectively describe both early-time inflation and late-time cosmic acceleration. Several authors have 
investigated several cosmological consequences using power law form of $f(Q)$. In the framework of the power-law model $f(Q)=\alpha Q^{n+1} +\beta$, Koussour and Bennai \cite{Koussour2022} examined the anisotropic cosmic evolution, showing congruence with quintessence-like behaviour and late-time acceleration. In another work Ghosh and Chattopadhyay \cite{Ghosh2024Inflation} studied the polynomial model of f(Q) gravity to investigate both bouncing and inflationary cosmology. Dynamical system analysis has been applied to scalar field cosmology within polynomial f(Q) gravity \cite{Ghosh2024Scalar}.\\

\subsection{Holographic fluid in $f(Q)$ gravity}
The concept of holography \cite{Susskind1995, Sutter2008 } has been utilized to constrain the magnitude of dark energy. The holographic dark energy model emerged as a useful method to consider the dark energy riddle among several dark energy models. In modern cosmology, the holographic principle has played an important role in providing a connection between quantum gravity and the large-scale structure of the Universe. Inspired by this idea, the Holographic Dark Energy model has been proposed as a viable candidate to explain late-time acceleration. On the other hand an alternative description of gravity on non-metricity is offered by $f(Q)$ Gravity, which is formulated within the symmetric teleparallel geometry. We will explore cosmological implications of the Holographic fluid in the context of $f(Q)$ gravity framework to study the early phase of the Universe as well as the late time of the Universe.

 In recent years, a significant number of studies have examined the implications of holographic dark energy within the context of context of $f(Q)$ gravity leading to a range of viable cosmological scenarios.
  The gravitational force $f(Q)$ has been thoroughly investigated by Saha and Rudra \cite{Saha2025} utilising holographic dark energy techniques. Additionally, they have demonstrated that their $f(Q)$ model satisfies the energy criteria and yields valuable, cosmologically applicable conclusions. Hatkar et al. \cite{Hatkar2024} discovered accurate solutions of the field equations under two alternative types of variable deceleration factors by examining the flat FRW universe with domain walls and holographic dark energy in the gravitational theory $f(Q)$. Their research thoroughly examined the behaviour of the physical parameters and demonstrated that the model is consistent with an expanding and speeding cosmos. Kalkan and Aktas \cite{Kalkan2026} investigate the cosmological dynamics of the Kaniadakis holographic dark energy model within the framework of $f(Q)$ Gravity in a flat FLRW universe. Using the Hubble horizon as the infrared cutoff, they have shown that the model exhibits behavior close to the Lambda-CDM Model and asymptotically approaches a de Sitter phase, indicating its viability in describing late-time cosmic acceleration. The dynamical study of holographic dark energy models has been analyzed within the framework of the FLRW Metric, where the gravitational Lagrangian is taken as an arbitrary function of the non-metricity scalar \cite{Shekh2021}. The author examined the evolution of key cosmological parameters, including the equation of state and stability parameters, and demonstrated that the model leads to a viable accelerating cosmological scenario. In another study, Dubey et al. \cite{Dubey2025} investigate the gravitational sector as the main driver of dark energy evolution within the framework of $f(Q)$ gravity, where the non-metricity scalar governs the dynamics, considering a model with both linear and nonlinear contributions. They also analyzed the model through statefinder and other diagnostic tools using initial conditions consistent with $\Lambda CDM$ and Planck 2018 data. The holographic dark energy entropies of Tsallis, Renyi, and Barrow were employed in $f(Q)$ theory by Saleem et al. \cite{Saleem2024}. Additionally, they confirmed that these entropy-based models are compatible with the Planck data confirming  the cosmological consistency.

 In this section we will discuss Barrow Holographic fluid in the background of $f(Q)$ gravity. Holographic dark energy density can be obtained by applying Barrow entropy in holographic framework and the energy density is of the form \cite{Saridakis2020}
 \begin{equation}\label{2eq1}
     \rho_{HDE}= C L^{\Delta-2}
 \end{equation}
  where $L$ the holographic horizon length and $C$ a parameter with dimensions $[L]^{-2-\Delta}$. For $\Delta = 0$, Barrow entropy reduces to the standard Bekenstein-Hawking form i.e Eq. (\ref{2eq1}) reduces to 
  
  \begin{equation} \label{2eq2}
     \rho_{HDE} = C L^{-2},
\end{equation}
where $C = 3c^2 M_p^2$ where $c^2$ is a dimensionless parameter of order unity appearing in holographic dark energy models \cite{Li2004}, and $M_p$ denotes the Planck mass. The Granda-Oliveros (GO) cutoff is given by \cite{Granda2008},
\begin{equation}\label{2eq3}
L^{-2}=\left(\gamma H^2+\delta \dot{H}\right).
\end{equation}
Now by applying  Eq.(\ref{2eq3}) in Eq.(\ref{2eq1}) the energy density is obtainde as
\begin{equation}\label{2eq4}
    \rho_{HDE}= 3 c {M_{p}}^2 \left(\gamma H^2+\delta \dot{H}\right)^{1-\frac{\Delta}{2}}.
\end{equation}

We consider a flat, homogeneous, and isotropic universe whose metric is given by 
 \begin{equation}\label{eq3a}
    ds^2=-dt^2+a^2(t)[dx^2+dy^2+dz^2]
\end{equation}
where $a$ is the scale factor. Imposing the metric (Eq. \ref{eq3a}) in the Einstein equation and considering the (00) component we obtain the First Friedmann equation as 
\begin{equation}\label{eq3b}
    \rho=3 H^2
\end{equation}
where $\rho$ is the total energy density of the universe.
In this study, we consider that the universe is filled with holographic fluid as well as the matter sector. The  Friedmann equations are written as 
\begin{equation}\label{2eq5}
    3H^2=\frac{1}{2f_{Q}}\left(-\rho+\frac{f}{2}\right)
\end{equation}
In Eq.(\ref{2eq5}), $\rho$ is the total energy density of matter and holographic fluid, i.e., $\rho=\rho_{HDE}+\rho_{m}$. We will use the Eq.(\ref{2eq5}) for the reconstruction of the Hubble parameter in the early and late time of the universe.

  \section{Inflationary Dynamics and Reconstruction scheme} \label{sec:infl}
In this section we consider that inflation is driven by holographic fluid in the background of $f(Q)$ gravity. 
  One of the important quantities for inflationary cosmology is the e-folding number, which is denoted by 
  \begin{equation}\label{3eq1}
      N=ln \frac{a_{f}}{a_{i}}
  \end{equation}
  The e-folding number $N$ denotes the number of Hubble times between two epochs corresponding to the scale factors $a_1$ and $a_2$. In the context of inflationary scenarios, the number of e-folds lies in the range $N \sim 45\text{-}60$ for  a successful resolution of the flatness and horizon problems. These estimations are mostly obtained in the context of inflationary models with a single scalar field. The precise value of the e-folding number, however, depends heavily on the overall equation-of-state parameter at the conclusion of inflation, and in some circumstances it may exceed 60 \cite{ Oikonomou2023}. 

Inflation refers to phase of accelerated expansion where the universe expands exponentially. During inflationary era the expansion is accelerated i.e. $\ddot{a}>0$. As a consequence the the comoving Hubble horizon shrinks i.e.
\begin{equation}\label{3eq2}
    \frac{d}{dt}\left(\frac{1}{aH}<0\right)
\end{equation}
Eq.(\ref{3eq2}) is mathematically equivalent to $\epsilon_1<1$ where 
\begin{equation}\label{3eq3}
    \epsilon_1=-\frac{\dot{H}}{H^2}.
\end{equation}
 $\epsilon_{1}$ is termed as first slow-roll parameters and the next slow roll parameters are defined as 
\begin{equation}\label{3eq4}
   \epsilon_{n+1}=\frac{\dot{\epsilon}_n}{H\epsilon_{n}}. 
\end{equation}
When inflation is taking place the slow roll indices \cite{Nojiri2017}:
\begin{equation}
    \epsilon_{1}, ~~ \epsilon_{2} <<1
\end{equation}
$\epsilon_{1}<<1$ ensures the occurence of the inflationary era and $\epsilon_{2}<<1$ ensures that inflation lasts for a sufficient amount of time \cite{Odintsov2023}.

  \subsection{Reconstruction of  the Hubble parameter}
To capture the inflationary dynamics within the framework of $f(Q)$ gravity, we reconstruct the Hubble parameter in terms of the e-folding number $N$, which provides a convenient description of cosmic evolution. 

 We consider the early time of the Universe in the presence of holographic fluid in the framework of $f(Q)$ gravity. In this regime, the matter contribution is considered to be negligible compared to the effective energy density arising from the modified gravity sector and the holographic dark energy component. Therefore, we consider matter energy density ($\rho_{m}$) to be zero, and under this assumption, the modified Friedmann equation (Eq.(\ref{2eq5})) reduces to 
 \begin{equation}\label{31eq1}
     6H^2 f_{Q}=-\rho_{HDE}+\frac{f}{2}
 \end{equation}
 We have already discussed the Barrow holographic energy density in the previous section. To proceed further, we adopt the form $\rho_{HDE}=3(\gamma H^2+\delta \dot{H})$ (considering $M_{p}$=1 and $c=1$). In the framework of $f(Q)$ gravity, we consider the non-linear form of $f(Q)$, i.e., $f(Q)=\zeta Q^n$ ($n>1$). In the FLRW background, the non-metricity scalar is given by $Q=6 H^2$. Therefore, the function $f(Q)$ and its derivative $f_{Q}$ takes the form
 \begin{equation}\label{31eq2}
  f(Q)=f(Q) = \zeta (6H^2)^n, ~~~f_{Q} = \frac{df}{dQ} = \zeta n Q^{n-1} = \zeta n (6H^2)^{n-1}.   
 \end{equation}
 Substituting these expressions along with $\rho_{HDE}$ in Eq.(\ref{31eq1}) and simplifying the equation we obtain
 \begin{equation}\label{31eq3}
     \zeta \left(n - \frac{1}{2}\right)(6H^2)^n = -3\left(\gamma H^2 + \delta \dot{H}\right).
 \end{equation}
 Finally, expressing the time derivative of the Hubble parameter $\dot{H}$ in terms of e-folding number using the expression $N=ln~a$ the above equation (Eq.(\ref{31eq3})) can be written as 
 \begin{equation}\label{31eq4}
     \frac{dH}{dN}= A H^{2n-1}-B H
 \end{equation}
where $A=(1-2n) \frac{\zeta}{\delta} 6^{n-1}$ and $B=\frac{\gamma}{\delta}$.
Solving the differential equation (Eq.(\ref{31eq4})) we obtain the reconstructed Hubble parameter as
\begin{equation}\label{31eq5}
    H=\left(\frac{A+e^{(N+c_{1})(2Bn-2B)}}{B}\right)^{\frac{1}{2(1-n)}}
\end{equation}
Since We are taking $n>1$, the coefficient in the numerator has to be positive to ensure the expressions remain positive and physically meaningful.  
We now use this reconstructed Hubble parameter $H(N)$ we proceed to evaluate the inflationary observables like slow-roll parameters, scalar spectral index , tensor to scalar ratio.

\subsection{Inflationary Observable}
 The main goal of gravitational theories in accurately describing cosmological inflation is to provide predictions that are consistent with observational data. In order to do this, it is necessary to look at the following theoretical predictions of the suggested gravity models for the inflationary observables \cite{Martin2014, Liddle2000, Schwarz2001}. 
  The scalar spectral index is given by 
  \begin{equation}\label{32eq1}
      n_{s}= 1+ \frac{d ~ln~ \mathcal{P_{S}}^2}{d~ln ~k}=1-6\epsilon+2\eta
  \end{equation}
  Tensor to scalar ratio is given by
  \begin{equation}\label{32eq2}
      r=\frac{\mathcal{P_{T}}^2(k)}{\mathcal{P_{S}}^2(k)}=16 \epsilon=16 \epsilon_{1}
  \end{equation}
  $\mathcal{P_{S}}$ , $\mathcal{P_{T}}$ are respectively the dimensionless power spectrum for scalar perturbations and tensor perturbations, and $k = aH$. In Eq.(\ref{32eq1}), $\eta=\epsilon_{1}-\epsilon_{2}$. 

  In previous section we have reconstructed the Hubble parameter (Eq.(\ref{31eq5}) in term of the e-folding number. We will use this reconstructed Hubble parameter to obtain the inflationary observables. Using Eq.(\ref{3eq3}) and Eq.(\ref{3eq4})   We obtain slow roll parameters in terms of e-folding number:
 
\begin{equation}\label{32eq3}
\epsilon_{1}=\frac{6 e^{\frac{2 (n-1) (c_1 + N) \gamma}{\delta}} \gamma}
{-6^{n} (2n-1)\zeta + 6 e^{\frac{2 (n-1) (c_1 + N) \gamma}{\delta}} \delta}
  \end{equation} 

  \begin{equation}\label{32eq4}
     \epsilon_{2}=
-\frac{\gamma \left(6^{n}(1-3n+2n^{2})\zeta
+6 e^{\frac{2(n-1)(c_{1}+N)\gamma}{\delta}}\delta\right)}
{\delta \left(-6^{n}(2n-1)\zeta
+6 e^{\frac{2(n-1)(c_{1}+N)\gamma}{\delta}}\delta\right)}.
  \end{equation}
Implementing the equations (Eq.(\ref{32eq3}), Eq.(\ref{32eq4})) in Eq.(\ref{32eq1}) and Eq.(\ref{32eq2}) we obtain the scalar spectral index  and tensor to scalar ratio as:
\begin{equation}\label{32eq5}
    n_{s}=
\frac{6^{n}(2n-1)\zeta\left(2(n-1)\gamma-\delta\right)
-6e^{\frac{2(n-1)(c_{1}+N)\gamma}{\delta}}(2\gamma-\delta)\delta}
{\delta\left(-6^{n}(2n-1)\zeta
+6e^{\frac{2(n-1)(c_{1}+N)\gamma}{\delta}}\delta\right)} 
\end{equation}

  \begin{equation}\label{32eq6}
      r =
\frac{96\,e^{\frac{2(n-1)(c_{1}+N)\gamma}{\delta}}\gamma}
{-6^{n}(2n-1)\zeta
+6\,e^{\frac{2(n-1)(c_{1}+N)\gamma}{\delta}}\delta}
  \end{equation}
  Thus we obtain explicit form of $n_{s}$, $r$, $\epsilon$, $\eta$ in terms of the model parameters $\zeta$, $\delta$, $\gamma$, $n$. In this way reconstruction of the Hubble parameter $H(N)$ in presence of holographic fluid within the framework $f(Q)$ gives path to study inflationary predictions.

  \subsection{Comparison with Observational Data}
   It is known from the Planck collaboration, the latest constraints on the scalar spectral index ($n_{s}$) and the tensor-to-scalar ratio ($r$) are \cite{Planck2018}:
\begin{equation}
    n_S = 0.9649 \pm 0.0042 \,\,\text{at}\,\, 68\%\,\,\text{CL}, \quad r < 0.056 \,\,\text{at}\,\, 95\%\,\,\text{CL} \,\,\text{.}
\end{equation}\\
In this work inflationary observables has evaluated for e-folding numbers $N = \{50, 60, 70\}$, which correspond to the typical range required to solve the horizon and flatness problems. The numerical results are represented in Table~\ref{tab:inflation}. It is clear from the table that the theoretical predictions of the proposed Holographic $f(Q)$ gravity model for $n_S$ and $r$ are in good agreement with the latest Planck observational data.

\begin{table}[htbp]
\centering
\caption{ Scalar spectral index $n_S$ and tensor-to-scalar ratio  $r$ for different e-folding numbers $N$.}
\label{tab:inflation}
\begin{tabular}{|c|c|c|}
\hline
\textbf{$N$} & \textbf{$n_S$} & \textbf{$r$} \\ \hline
50 & 0.966667 & 2.80 $\times 10^-6$ \\ \hline
60 & 0.966668 & 3.91 $\times 10^-6$\\ \hline
70 & 0.966668 & 5.45 $\times 10^-6 $\\ \hline
\end{tabular}
\end{table}

\begin{figure}[htbp]
\begin{center}
    \includegraphics[scale=0.40]{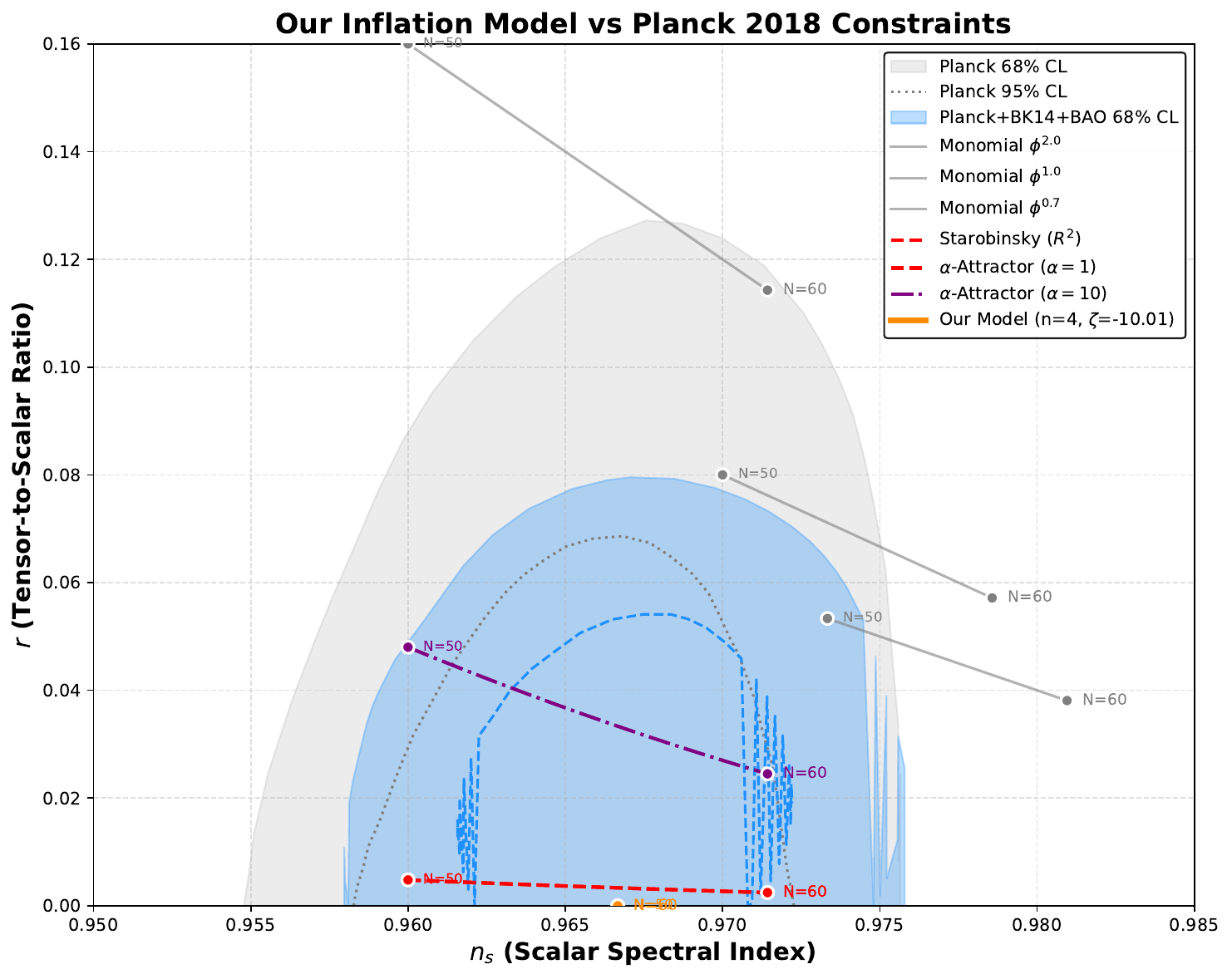}
\caption{spectral index versus tensor to scalar ratio}
\label{f1}
\end{center}
\end{figure}

Fig.\ref{f1} represents the scalar spectral index $n_S$ vs tensor-to-scalar ratio  $r$. In this figure we compare our model with Planck 2018 constraints and several well-known inflationary scenarios: 

\begin{itemize}
    \item \textbf{Starobinsky ($R^2$) model}: In this benchmark inflationary model based on $R + R^2$ gravity, predicting scalar spectral index and tensor to scalar ratio to be $n_S \approx 0.965$ and $r \approx 0.004$ for $N=60$.
    
    \item \textbf{$\alpha$-attractors}: A class of inflationary models where the predictions depend on the parameter $\alpha$. For $\alpha = 1$ and $\alpha = 10$, these models occupy distinct regions in the $n_S$-$r$ plane. For $\alpha=1$ the model coincides with the Starobinsky inflation model.
    
    \item \textbf{Monomial models}: In monomial models the scalar field (inflaton) evolves under a power law potential where $V(\phi) \propto \phi^p$ with $p = 0.7, 1.0, 2.0$. These typically predict larger tensor-to-scalar ratios.
    
    \item \textbf{Our  model}: It is apparent from Fig.~\ref{f1} that the predictions of our model for $N=50$ and $N=60$ lie well within the Planck 68\% confidence region, with a lower tensor-to-scalar ratio ($r \sim 10^{-6}$), consistent with the non-detection of primordial B-modes.
\end{itemize}
In Fig.\ref{f1} we have presented the behavior of the scalar spectral index $n_S$ and tensor-to-scalar ratio $r$ derived from the holographic $f(Q)$ model. The orange line as indicated in the figure corresponds to $n_{s}$- $r$ trajectory  for our model with $N=50$ and $N=60$. In this case it is very close to Starbinsky model and fully contained within Planck 2018 confidence contour. It is evident from the figure that our model predicts a very low value of $r$ and maintained acceptable values of $n_{s}$. The small value of $r$ indicates that the model presented here subpresses primordial tensor perturbations and the model appears to be consistent with the  CMB observational constraint.

\section{Late Time Cosmology with matter sector} \label{sec:matter}
A number of studies have sought to construct a unified description connecting the early-time inflationary phase and the late-time accelerated expansion within a single theoretical framework\cite{Nojiri2024a, Nojiri2024b, Allemandi2005, Nojiri2004,Nojiri2011, Nojiri2017, Nojiri2003, Nojiri2006, Cognola2008,Nojiri2007b}. Such unification is very attractive as it offers a consistent understanding of the expansion of the universe without invoking distinct mechanisms for different cosmological eras. In this direction, modified gravity theories have played a crucial role in reproducing both the inflationary phase and present cosmic acceleration. In this section, we investigate the late-time cosmological implications of the holographic $f(Q)$ gravity model in the presence of the matter sector. Contrary to various earlier studies, where the inflationary and late-time acceleration are treated separately, the current study considers both epochs a single holographic $f(Q)$ framework. Since the nonlinear field equation obtained for $n>1$ does not admit an analytical solution for $H(a)$, we employ the CPL parametrization to connect the theoretical model with observational cosmology. The model parameters are further constrained using combined CC+BAO datasets through MCMC analysis in order to examine the observational viability of the proposed scenario at late times.

  \subsection{Matter Dark Energy Coupling}
  In standard cosmology dark energy and dark matter evolve independently, with their energy–momentum tensors conserved separately. Since we assume that the universe is filled with both pressureless matter and holographic fluid, the Eq.(\ref{2eq5}) receives contributions from both sectors. Therefore the Eq.~(\ref{2eq5}) can be written as
\begin{equation}\label{41eq1}
6 \zeta n H^2 Q^{\,n-1}
= -\rho_m - \rho_{\mathrm{HDE}} + \frac{\zeta Q^n}{2}.
\end{equation}
From conservation equation of matter we obtain $\rho_m = \rho_{m0} a^{-3}$. Now incorporating $\rho_m$ and $\rho_{HDE}$ in equation Eq.(\ref{41eq1}) and performing  algebraic simplifications, we obtain a nonlinear differential equation governing the dynamics of the Hubble parameter as
\begin{equation}\label{41eq2}
P H^2 + S \dot{H} + R H^{2n} = \frac{\rho_{m0}}{a^3},
\end{equation}
where the coefficients are defined as
\begin{equation}
P = -3\gamma, \qquad 
S = -3\delta, \qquad 
R = \frac{6^n(1-2n)}{2}\,\zeta.
\end{equation}
The term $H^{2n}$ present in the Eq.(\ref{41eq2}) is highly non linear term because  we take $n\neq1$. Now the interplay between the parameters like ${\zeta}$, $n$, $\delta$, $\gamma$ will decide whether the universe will accelerated expansion or possible transition between different phases of cosmic acceleration.

  \subsection{CPL parameterization for late time evolution}
 % In $f(Q)$ gravity, the energy density is of the form
%\begin{equation}\label{4eq1}
 % 3H^2=\frac{1}{2f_{Q}}(-\rho+\frac{f}{2})  
%\end{equation}  
%For this work, we choose the polynomial model of $f(Q)$, i.e., $f(Q)=\zeta Q^n$. The modified Friedmann equation takes the form 
%\begin{equation}\label{4eq2}
    %6 \zeta n H^2 Q^{n-1}= (-\rho_{m}-\rho_{HDE}+\frac{\zeta Q^n}{2})
%\end{equation}.
 Since the cosmological field equations do not admit analytical solutions we consider a phenomenological anstz in the form of CPL parameterization. To be more specific
the differential equation in Eq.(\ref{41eq2}) can not be solved in terms of $H(a)$ because we choose $n>1$. Therefore, instead of solving $H(a)$ directly, we will use the parametric form of $H(a)$ based on observation. Next, we will make this ansatz to satisfy Eq.(\ref{41eq2}) at the present epoch ($a=1$) up to higher order derivatives. In this manner we will try to find the relation between ${P,S,R,n}$ with the ansatz parameters. 

The Chevallier-Polarski-Linder (CPL) parametrization  is widely used in late-time cosmology. Since our model contains four parameters $(P,S,R,n)$ to be constrained observationally, we require an ansatz involving four observable parameters. The CPL parametrization satisfies this requirement, which is the reason for choosing the CPL parametric form in the present model. Moreover, the CPL parametrization allows the equation of state parameter to evolve with cosmic time. It expresses equation of state in terms of scale factor as \cite{Chevallier2001,Linder2003, Dantas2020, Scherrer2015}
\begin{equation}\label{4eq4}
    w(a)=w_{0}+w_{b} (-a+1)
\end{equation}
where $w_{0}$ denotes the present value and $w_{b}$ measures the evolution of dark energy with time. The CPL parametric form is

    \begin{equation} \label{4eq5}
    H^2(a) = H_0^2 \left[ \Omega_{m_0} a^{-3} + \Omega_{de_0} \, a^{-3(1 + w_{0} + w_{b})} \, e^{-3 w_{b} (1 - a)} \right]
\end{equation}
Here $\Omega_{m}+\Omega_{de}=1$
From Eq.(\ref{4eq5}) it is apparent that at present time ( $a=1$) $H^2=H^2_{0}$. 

For convenience we rewrite the Eq.(\ref{4eq5}) in the form 
\begin{equation}\label{4eq6}
   \frac{H^2}{H^2_{0}}= E^2(a)= \left[ \Omega_{m_0} a^{-3} + \Omega_{de_0} \, a^{-3(1 + w_{0} + w_{b})} \, e^{-3 w_{b} (1 - a)} \right]. 
\end{equation}
This equation satisfies the normalization condition $E=1$ at $a=1$. To extract constraints on the model parameters we introduce the differential operator
\begin{equation}\label{4eq7}
    \partial_a=a\frac{d}{da}
\end{equation}
Now we apply $\partial_a$ in Eq.(\ref{4eq6}) and in the differential  Eq.(\ref{41eq2}) and putting $a=1$ we obtain 
\begin{equation}\label{4eq8}
    -3P A_1+\frac{9}{2} S B_2-3 nR H_{0}^{2n-2}=-9 \Omega_{m_{0}}
\end{equation}
where $B_{1}= 1+w_{0} \Omega_{m_{0}}$

Again we apply the operator $\partial^2_a$ in Eq.(\ref{4eq6}) and in the relation $PH^2+S\dot{H}+RH^{2n}=\frac{\rho_{m_{0}}}{a^3}$ and then evaluating at $a=1$ we obtain 

\begin{equation} \label{4eq9}
    9 P B_1 + \frac{27}{2} S B_2 + 9 n R H_{0}^{2n-2}(2n - 1) B_1 = 27 \Omega_{m_{0}}
\end{equation}
where $B_{2}=3 \left[ 3 + (1 - \Omega_{m_{0}})\left( w_b + 6 w_0 + 3 w_0^2 \right) \right]$

Lastly, applying the operator $\partial^3_a$ to Eq.(\ref{4eq6}) and Eq.(\ref{41eq2}), and evaluating at $a = 1$, we obtain,
\begin{equation}\label{4eq10}
 -27 P B_1 - \frac{81}{2} S B_3 - 27 n R H_{0}^{2n-2}  B_1 (2n - 1)(2n - 2) = -81 \Omega_{m_{0}}
\end{equation}
where $B_{3}=3 \left[ -9 + (\Omega_{m_{0}} - 1)\left(9 w_0^3 + 27 w_0^2 + 27 w_0 + 9 w_0 w_b + 8 w_b \right) \right]$.

We need one more relation for evaluating four model parameters. Therefore taking $a=1$ in the relation $PH^2+S\dot{H}+RH^{2n}=\frac{\rho_{m_{0}}}{a^3}  $ we obtain
\begin{equation}\label{4eq11}
P-\frac{3S}{2} B_{1}+RH^{2n-2}_{0}= 3\Omega_{m_{0}}
\end{equation}
We can isolate $P$ from Eq.(\ref{4eq11}) and $P$ will take the form

\begin{equation} \label{4eq12}
    P = 3 \Omega_{m_{0}} + \frac{3}{2} S B_1- R H_{0}^{2n-2} 
\end{equation}
Substituting $P$ in Eq.(\ref{4eq8}) we obtain the algebric expression for $S$ which is of the form
\begin{equation}\label{4eq13}
     S = \frac{2 \Omega_{m_{0}} (A_1 - 1) - \frac{2}{3} R H_{0}^{2n-2}  B_1 (1 - n)}{B_2 - B_1^2}
\end{equation}
Now sustituting the expressions for $P$ and $S$ in Eq.(\ref{4eq9}) we can find out the expression for $R H_{0}^{2n-2} $ which is of the form
\begin{equation}\label{4eq14}
    R H_{0}^{2n-2} = \frac{27 \Omega_{m_{0}} (1 - B_1) - \frac{27}{2} S (B_3 + B_1^2)}{9 n B_1 (2n - 1) - 9 B_1}
\end{equation}
Finally the value of $n$ can be obtain by substituting the expressions of $P$, $S$ and $R H_{0}^{2n-2}$ in Eq.(\ref{4eq10}).

 \subsection{Observational Constraints at Late time }
  \subsubsection{Cosmic Chronometers (CC)}

In this article, we use 31 independent measurements of the Hubble parameter $H(z)$ obtained from the cosmic chronometer method \cite{Moresco2016 ,Solanki/2021}. These measurements are derived from the differential ages of massive, passively evolving galaxies, and this provides model-independent estimates of the expansion rate through:
\begin{equation}
    H(z) = -\frac{1}{1+z}\,\frac{dz}{dt}.
\end{equation}
The data used in this study consists of 31 points spanning the redshift range $0.07 \leq z \leq 2.41$ \cite{Solanki/2022}. 
\subsubsection{Baryon Acoustic Oscillations (BAO)}

Along with CC data we also include 26 BAO observations from various galaxy surveys and Lyman-$\zeta$ forest measurements \cite{Solanki/2021}. These data points provide standard ruler measurements that constrain the expansion history and angular diameter distance. Our analysis uses 26 data points over the redshift range $0.10 \leq z \leq 2.36$, bringing the total number of observational data points to 57.

In this article, we combine CC and BAO data because they probe complementary aspects of the expansion history. That is, CC measurements provide direct, differential constraints on $H(z)$, while BAO provides integrated-distance information. This complementarity helps break degeneracies between $H_0$, $\Omega_m$, and the dark energy parameters \cite{Moresco2016}, leading to tighter and more robust constraints than either dataset alone.

\subsection{Statistical Methodology}\label{subsec:stats}

\subsubsection{Chi-Square Analysis}

For the CC data, the standard chi-square statistic that is used for observational data fitting or parameter estimation in cosmology as:
\begin{equation}
    \chi^2_{\text{CC}} = \sum_{k=1}^{31}\frac{\left[H_{\text{th}}(z_k)-H_{\text{obs},k}\right]^2}{\sigma_{H,k}^2},
\end{equation}
where $H_{\text{th}}(z_k)$ is the theoretical prediction, $H_{\text{obs},k}$ is the observed value, and $\sigma_{H,k}$ is the measurement uncertainty. Note that these data points are uncorrelataed 

For the BAO data, we account for correlated errors using the covariance matrix $\mathbf{C}$:
\begin{equation}
    \chi^2_{\text{BAO}} = (\mathbf{A}_{\text{obs}} - \mathbf{A}_{\text{th}})^T \mathbf{C}^{-1} (\mathbf{A}_{\text{obs}} - \mathbf{A}_{\text{th}}).
\end{equation}

The total chi-square is given by:
\begin{equation}
    \chi^2_{\text{total}} = \chi^2_{\text{CC}} + \chi^2_{\text{BAO}}.
\end{equation}

\subsubsection{MCMC Sampling}

We perform parameter estimation using Markov Chain Monte Carlo (MCMC) sampling with the \texttt{emcee} package \cite{ForemanMackey2013}. The posterior probability distribution is:
\begin{equation}
    P(\mathbf{p}|\text{data}) \propto \exp\left(-\frac{\chi^2_{\text{total}}}{2}\right) \times \Pi(\mathbf{p}),
\end{equation}
where $\mathbf{p} = \{H_0, \Omega_{m0}, w_0, w_b, \}$ represents the free parameters and $\Pi(\mathbf{p})$ denotes uniform priors over physically motivated ranges.

\subsubsection{Parameter Constraints}

We report the best-fit values and confidence intervals at $1\sigma$ (68.3\%) and $2\sigma$ (95.4\%) confidence levels. The convergence of MCMC chains is verified using standard diagnostics, and marginalized posterior distributions are obtained for all parameters \cite{Solanki/2021}.

\begin{figure}[H]
\includegraphics[scale=0.60]{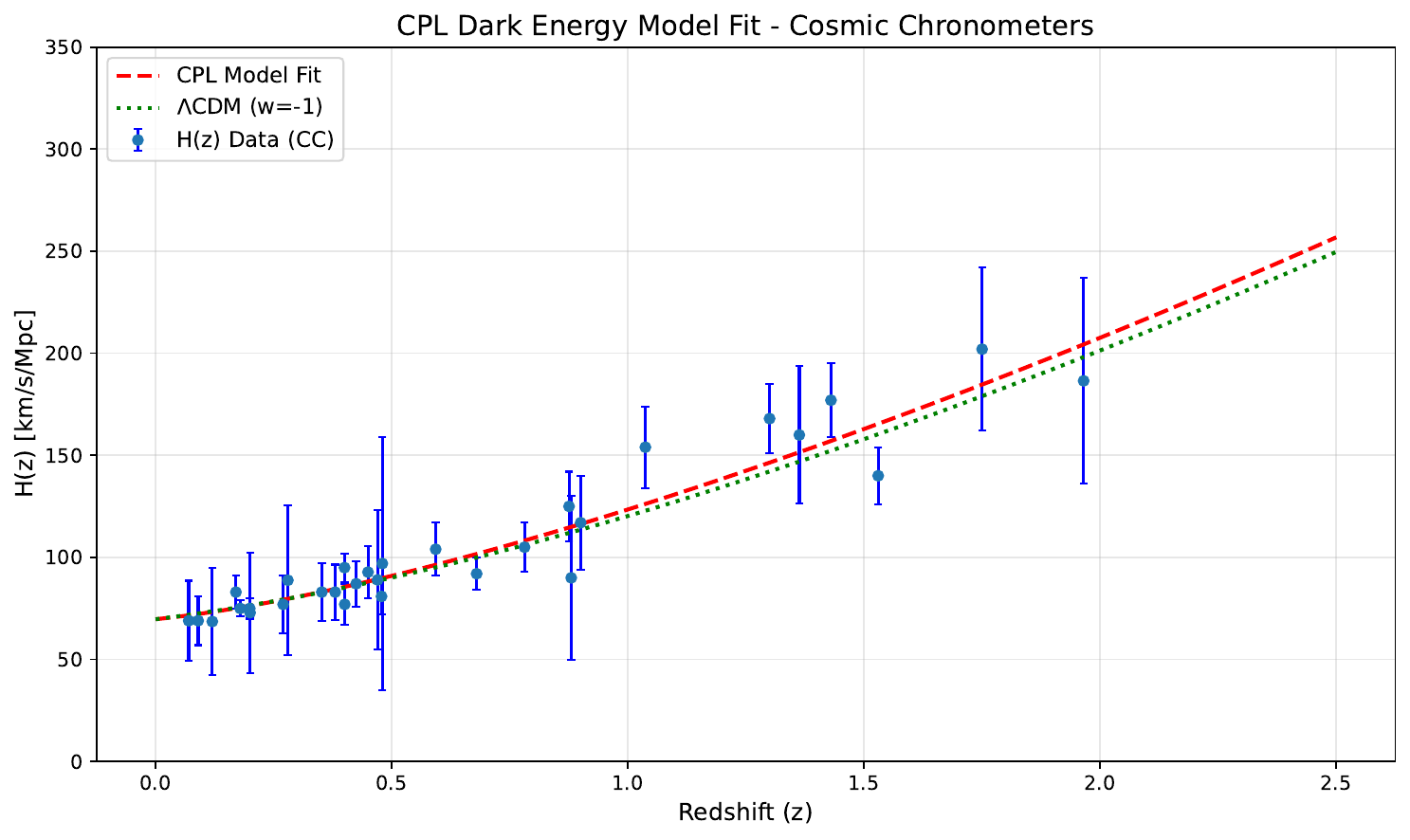}
\caption{This is the least square fitting for the holographic dark energy with CPL parametrization using Hubble data set (31 CC+26 BAO data points)}\label{f2}
\end{figure}

Figure~\ref{f2} shows the best-fit evolution of $H(z)$ for the CPL-parametrized holographic dark energy model using CC+BAO data (31 CC + 26 BAO points). The red-dashed line represents the CPL, and it is observed to be close to the observational data over the redshift range $0 \lesssim z \lesssim 2$ shown here. For $z \lesssim 0.6$  both CPL and $\Lambda$CDM ($w=-1$, green dotted) are almost coincident. However, at higher redshifts ($z \gtrsim 1$), the CPL model representing curve stays slightly above the $\Lambda$CDM. This implies that it predicts a slightly higher expansion rate. However, this deviation lies within the error bars, and hence, it is not statistically significant. Thus, the CPL model is comparable to $\Lambda$CDM at the background level, and the current dataset does not allow a clear discrimination between dynamical dark energy and the cosmological constant.

\begin{figure}[h!]
\begin{center}
   \includegraphics[scale=0.60]{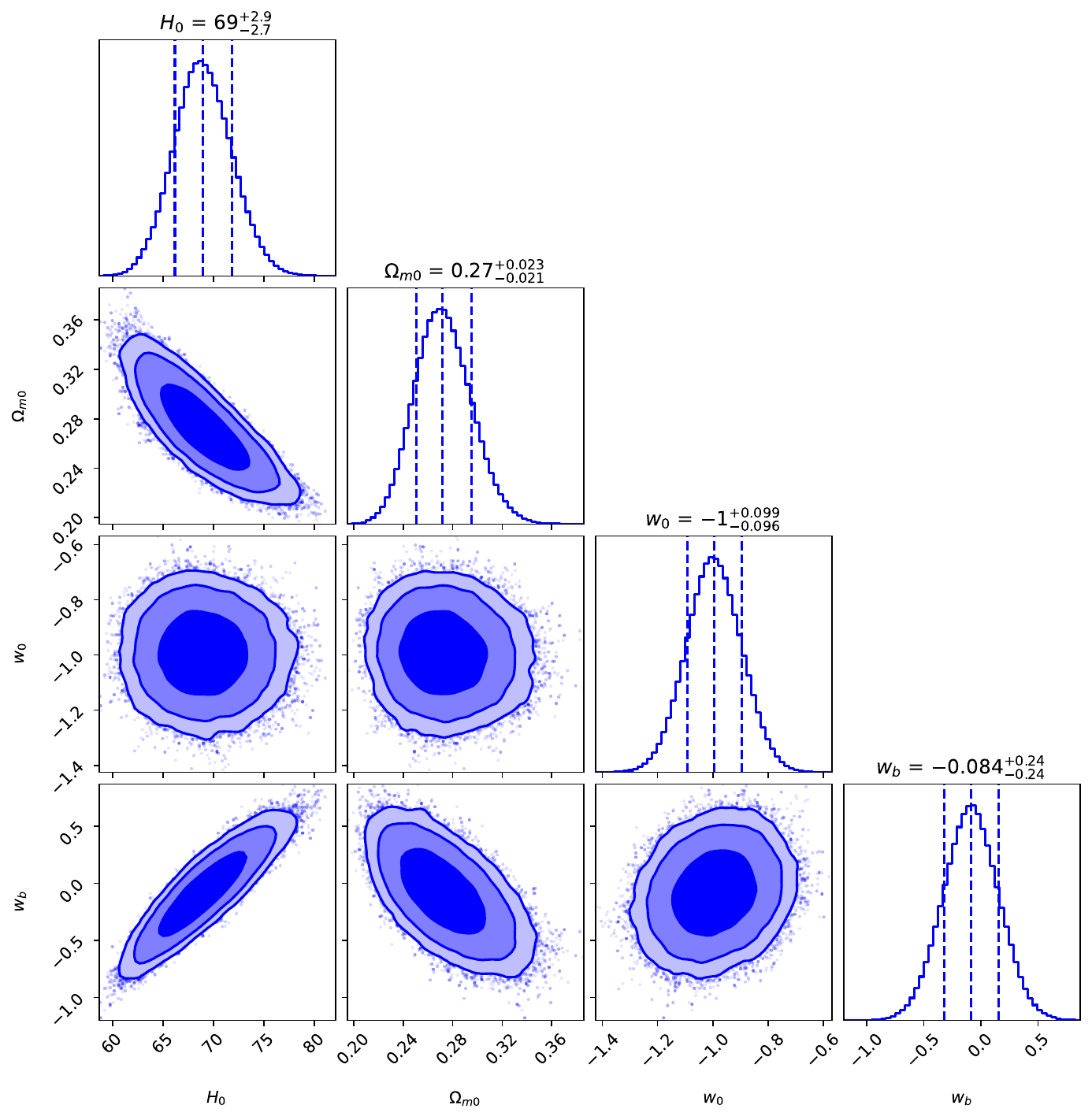}
\caption{This is the MCMC analysis for the holographic dark energy with CPL parametrization using Hubble data set (31 CC+26 BAO data points)}\label{f3} 
\end{center}
\end{figure}
In Fig.\ref{f3} we have presented MCMC corner plot for the Holographic Dark Energy model with CPL parametrization in $f(Q)$ gravity. We have constrained the model with a combined 31 CC + 26 BAO observational data points. The diagonal panels show posterior distributions, and the off-diagonal panels corresponds to confidence contours at $1\sigma$ , $2\sigma$ and $3\sigma$
levels corresponding to the parameters $H_{0}$, $\Omega_{m_0}$, $w_{0}$, $w_{b}$ . We have observed from the figures that the best-fit values obtained are approximately $H_0 = 69^{+2.9}_{-2.8}$ km/s/Mpc, $\Omega_m = 0.27^{+0.023}_{-0.021}$, $w_0 = -0.99^{+0.097}_{-0.099}$, and $w_b = -0.077^{+0.024}_{-0.024}$. The constraint value of $H_{0}$ is consistent with the $H_{0}$ constraint by Planck Data \cite{Planck2018}. The contour plot further reveals an anti-corelation between $H_{0}$ and $\Omega_{m_{0}}$. On the other $w_{b}$ has positive corelation with $H_{0}$ and negative corelation with $\Omega_{m_{0}}$. The constraint value of $w_0$ is close to cosmological constant and value of $w_b$ suggest mild effect of the parametrization on the dynamical evolution for the model under consideration. Finally Fig.\ref{f3}  shows that the confidence regions are compact and the constraint parameter values are consistent with observation indicating CPL holographic $f(Q)$ model remains well consistent with late time cosmological observation and an accelerated expansion of the universe. 

\subsection{Model comparison via AIC and BIC}
As given a model and data one can always over-fit by increasing the number of free parameters so, the Akaike (AIC) and Bayesian (BIC) Information Criteria \cite{Akaike/1974,Schwarz/1978} quantify the trade-off between fit quality and model complexity. Following the idea \cite{Liddle/2007}, we compute both from our MCMC $\chi^2_{\rm min}$ values and compare models via $\Delta{\rm IC} = {\rm IC}_{\rm CPL} - {\rm IC}_{\Lambda{\rm CDM}}$, with standard thresholds which are given as \cite{Burnham/2002,Kass/1995}: $\Delta{\rm AIC} < 2$ (substantial support for CPL), $2$--$4$ (positive against), $4$--$7$ (moderate against), $\geq 7$ (strong against). For BIC, the usual benchmarks are that $\Delta{\rm BIC} \sim 2$, $6$, or $10$ signals weak, moderate, or strong evidence against the CPL model, respectively.

For a model with $d$ free parameters fitted to $N$ data points,
\begin{equation}
{\rm AIC} = \chi^2_{\rm min} + 2d\,,\qquad
{\rm BIC} = \chi^2_{\rm min} + d\,\ln N\,.
\label{eq:aic_bic}
\end{equation}
The CPL parameterization uses $d=4$ parameters ($H_0$, $\Omega_{m_0}$, $w_0$, $w_b$) fitted to $N=57$ CC+BAO points; $\Lambda$CDM uses $d=2$. Priors and best-fit values are listed in Table~\ref{tab:priors_cpl}; information criteria are in Table~\ref{tab:aic_bic_cpl}.

\begin{table}[ht]
\centering
\caption{Gaussian priors and best-fit CPL parameters (68\% credible intervals) from joint CC+BAO analysis.}
\label{tab:priors_cpl}
\begin{tabular}{lcc}
\hline
Parameter & Prior & Best-fit $\pm 1\sigma$ \\
\hline
$H_0$ [km s$^{-1}$ Mpc$^{-1}$] & $70.0 \pm 5.0$ & $69.04 \pm 2.86$ \\
$\Omega_m$ & $0.20 \pm 0.10$ & $0.27 \pm 0.02$ \\
$w_0$ & $-1.02 \pm 0.10$ & $-0.99 \pm 0.10$ \\
$w_b$ & $0.50 \pm 0.50$ & $-0.08 \pm 0.24$ \\
\hline
\end{tabular}
\end{table}

\begin{table}[ht]
\centering
\caption{AIC/BIC comparison for CPL vs $\Lambda$CDM ($N=57$). Positive $\Delta$IC favours $\Lambda$CDM.}
\label{tab:aic_bic_cpl}
\begin{tabular}{lcccc}
\hline
Model & $d$ & $\chi^2_{\min}$ & AIC & BIC \\
\hline
CPL          & 4 & 35.23 & 43.23 & 51.40 \\
$\Lambda$CDM & 2 & 35.51 & 39.51 & 43.59 \\
\hline
$\Delta$IC   &   & $-0.28$ & $+3.72$ & $+7.81$ \\
\hline
\end{tabular}
\end{table}

We find that $\Delta{\rm AIC} = +3.72$ and $\Delta{\rm BIC} = +7.81$, indicating positive-to-strong evidence for $\Lambda$CDM. Which actually aligns with our constraints: $w_0 = -0.99 \pm 0.10$ and $w_b = -0.08 \pm 0.24$ are consistent with $\Lambda$CDM ($w_0=-1$, $w_b=0$) at $1\sigma$, so the extra CPL parameters do not yield a sufficient $\chi^2$ improvement to justify their complexity.

The high value of $\Delta{\rm BIC}$ also highlights a limitation of CPL as a phenomenological expansion, it lacks a fundamental theoretical anchor. Progress requires either a physically motivated parameterization or tighter data on $w_b$. Our coupled $f(Q)$ framework (Section~4) addresses the former. Mapping the CPL best-fit values onto the model parameters $P$, $S$, $R$, $n$ via Eqs.~(\ref{4eq8})--(\ref{4eq14}) yields rough estimates: $n \approx 1.02$, $P \approx 0.8$, $S \approx -0.15$, $R H_0^{2n-2} \approx 0.8$. Translating to fundamental couplings, $\gamma = -P/3 \approx -0.26$, $\delta = -S/3 \approx +0.05$, $\zeta \approx -0.2$ (large uncertainty from $1-2n \approx -1$). All of the above results are compatible with the GR limit ($n=1$, $\gamma=\delta=0$), even though we have taken $f(Q)$ gravity, this reinforces that current data prefer a model very close to $\Lambda$CDM. The broad uncertainties on $\zeta$ and the couplings reflect the limitation of background expansion alone. In order to constrain these further, growth-of-structure or high-$z$ distance data would be needed.

While current CC+BAO data favor $\Lambda$CDM, our coupled $f(Q)$ holographic dark energy framework remains viable, and future growth-of-structure measurements ($\sigma_8$, $f\sigma_8$) will be essential to probe its subtle deviations from the standard paradigm.

\section{Concluding Remarks} \label{sec:conclusion}

In this work, we have investigated a unified cosmological scenario in the framework of holographic $f(Q)$ gravity, where both the early inflationary era and the late-time accelerated expansion of the Universe have been studied in a single theoretical setup. We have considered the power-law form $f(Q)=\zeta Q^{n}$ and reconstructed the Hubble parameter in presence of holographic fluid. The reconstructed form of the Hubble parameter given in Eq.~\eqref{31eq5} provides a suitable description for inflationary dynamics.

Using the reconstructed Hubble parameter, we have evaluated the slow-roll parameters, scalar spectral index $n_s$, and tensor-to-scalar ratio $r$ through Eqs.~\eqref{32eq3}--\eqref{32eq6}. The numerical values obtained are presented in Table~\ref{tab:inflation} and the table shows that the model predictions are well consistent with the latest Planck observational data. In particular, the obtained value of the tensor-to-scalar ratio remains very small, which indicates suppression of primordial tensor perturbations. Furthermore, Fig.~\ref{f1} shows the behaviour of the scalar spectral index versus tensor-to-scalar ratio plane for the model under consideration. It is observed that the trajectory of our model lies well within the Planck 2018 confidence contour and remains very close to that of the Starobinsky inflationary model. Therefore, it may be stated that the present holographic $f(Q)$ gravity model appears to be observationally viable from the inflationary point of view.

In order to study the late-time cosmological evolution, we have incorporated the matter sector along with the holographic dark energy component. Since the nonlinear differential equation in Eq.~\eqref{41eq2} can not be solved analytically for $n>1$, we have adopted the CPL parametrization given in Eq.~\eqref{4eq5}. Using the observational Hubble data sets consisting of 31 CC and 26 BAO data points, we have constrained the model parameters through MCMC analysis. The least square fitting that is pictorially presented in Fig.~\ref{f2} demonstrates that the CPL holographic model remains very close to the standard $\Lambda$CDM cosmology at low redshift, while a small deviation appears at the region of higher redshift. However, this deviation still remains inside the observational error bars. Moreover, the MCMC contour plot shown in Fig.~\ref{f3} indicates that the obtained best-fit values of $H_0$, $\Omega_{m0}$, $w_0$, and $w_b$ are consistent with recent observational constraints. The confidence contours are also compact, indicating stability and observational consistency of the model. We also checked whether the extra CPL parameters are really needed via AIC and BIC criteria. The information criteria give $\Delta{\rm AIC}=+3.72$ and $\Delta{\rm BIC}=+7.81$ relative to $\Lambda$CDM, so the data mildly prefer the simpler model, which makes sense given that $w_0$ and $w_b$ sit right on the $\Lambda$CDM values within $1\sigma$. Pushing further, if we map these constraints onto the coupled $f(Q)$ parameters of Section~4, we get $n\approx1.02$ and couplings $\gamma,\delta,\zeta$ all compatible with zero (though $\zeta$ is poorly constrained). In short, current CC+BAO data do not demand deviations from $\Lambda$CDM or strong matter and dark energy coupling; any such effects must be subtle and would likely require growth or high-$z$ data to uncover.

The present work therefore highlights the holographic $f(Q)$ gravity as a unified cosmological framework unifying the early inflationary epoch and the late-time accelerated epoch within a single cosmological setup. Unlike plethora of existing studies, the current analysis combines inflationary reconstruction, observational inflation constraints, CPL-based late-time parametrization, and CC+BAO observational analysis in the same framework. The consistency of the model with both Planck inflationary constraints and late-time observational datasets indicates that the proposed scenario remains a viable alternative for explaining the complete cosmological evolution of the universe. Therefore, from the above analysis, we conclude that the holographic $f(Q)$ gravity model can successfully provide a unified description of inflation and late-time cosmic acceleration. The inflationary observables are found to be consistent with the latest Planck constraints, while the late-time analysis using the CC+BAO datasets supports a viable accelerating cosmology close to the $\Lambda$CDM scenario. Hence, the proposed model remains cosmologically viable and may serve as an alternative framework for explaining both the early- and late-time evolution of the Universe. Future investigations may include perturbation analysis, stability analysis, and thermodynamical aspects of the present model in greater detail.

\section*{Acknowledgment}
 The author Surajit Chattopadhyay acknowledges the visiting associateship of the Inter-University Centre for Astronomy and Astrophysics (IUCAA), Pune, India.

\section*{Data availability statement}
The observational datasets used in this work, namely the Cosmic Chronometer (CC) and Baryon Acoustic Oscillation (BAO) datasets, are publicly available from the corresponding published literature cited in the References. No new datasets were generated during the current study.

\section*{Declaration of generative AI and AI-assisted technologies in the writing process}
During the preparation of this manuscript, the authors used Grammarly and QuillBot for language improvement and grammar correction. After using these tools, the authors reviewed and edited the content as needed and take full responsibility for the content of the published article.

% -------------------------------------------------------------------------
% Suggested compact illustrative figure for the conclusion (insert where desired)
% -------------------------------------------------------------------------

%\section{Acknowledgement}

\end{document}